\documentclass[aps,prb,twocolumn,showpacs,superscriptaddress]{revtex4}

\usepackage{amsmath}
\usepackage{epsf}
\usepackage{epsfig}

\begin{document}

\title{Hydrogen storage of calcium atoms adsorbed on graphene: First-principles plane wave calculations}

\author{C. Ataca}
\affiliation{Department of Physics, Bilkent University, Ankara
06800, Turkey}\affiliation{UNAM-Institute of Materials Science and
Nanotechnology, Bilkent University, Ankara 06800, Turkey}
\author{E. Akt\"{u}rk}
\affiliation{UNAM-Institute of Materials Science
and Nanotechnology, Bilkent University, Ankara 06800, Turkey}
\author{S. Ciraci} \email{ciraci@fen.bilkent.edu.tr}
\affiliation{Department of Physics, Bilkent University, Ankara
06800, Turkey}
\affiliation{UNAM-Institute of Materials Science
and Nanotechnology, Bilkent University, Ankara 06800, Turkey}

\date{\today}

\begin{abstract}
Based on the first-principles plane wave calculations, we showed that Ca adsorbed on graphene can serve as a high-capacity hydrogen storage medium, which can be recycled by operations at room temperature. Ca is chemisorbed by donating part of its $4s$-charge to the empty $\pi^*$-band of graphene. At the end adsorbed Ca atom becomes positively charged and the semi-metallic graphene change into a metallic state. While each of adsorbed Ca atoms forming the (4$\times$4) pattern on the graphene can absorb up to five H$_2$ molecules, hydrogen storage capacity can be increased to 8.4 wt \% by adsorbing Ca to both sides of graphene and by increasing the coverage to form the (2$\times$2) pattern. Clustering of Ca atoms is hindered by the repulsive Coulomb interaction between charged Ca atoms.
\end{abstract}


\maketitle

In order to develop an efficient medium of hydrogen storage, carbon based nanostructures functionalized by transition metal atoms have been a subject of active study\cite{zhao,dag,yildirim,durgun,ataca}. Recently, Yoon \emph{et al.}\cite{yoon} have demonstrated that covering the surface of C$_{60}$ with $32$ Ca atoms can store 8.4 wt \% hydrogen. Their result, which is crucial for safe and efficient hydrogen storage\cite{coontz}, inspired us to consider graphene as the substrate material for Ca atoms. Graphene is precursor to C$_{60}$ and carbon nanotubes, but being a single atomic plane of graphite its both sides may be suitable for the adsorption of Ca atoms. Graphene by itself has been synthesized
showing unusual electronic and magnetic properties\cite{novo}.

In this paper, we showed that Ca atoms, in fact, can be
bound on both sides of graphene plane and each Ca atom absorbing
four H$_{2}$ results in a medium of high-capacity hydrogen storage of $8.4$
wt \%. In the present case the binding energy of the fourth
H$_{2}$ absorbed by Ca atom is still significant and is $\sim$300
meV. While each Ca atom donates part of its charge to the graphene
layer, graphene, by itself, having Fermi surface consisting of six points
at the corners of the hexagonal Brilloiun zone, is metallized.
These results are obtained from our study based on first-principles calculations\cite{dft}.

We first consider the adsorption of a single Ca on the graphene as
the substrate material. This is modeled by one Ca atom adsorbed on the
hollow site (namely H1 site above the center of hexagon) for each
(4$\times$4) cell of graphene (namely one Ca atom for every $32$ carbon
atoms). The Ca-Ca interaction is indeed negligible owing to large distance of $\sim 9.84$ \AA~ between them. A chemical bonding occurs between Ca and C atoms with a
binding energy of $0.99$ eV and Ca+graphene distance
of $2.10$ \AA. Similar to the bonding mechanism of Ca on C$_{60}$, Ca
atom donates part of its charge from $4s$-orbital to the $\pi^*$-bands of graphene. Due to the formation of an electric field between
Ca atom and the graphene layer, part of this charge is then back-donated\cite{yoon}
to the unoccupied $3d$-orbitals of Ca through their hybridization with
$\pi^*$-states. The resulting positive charge of Ca atom is calculated
to be $\sim 0.96 $ electrons\cite{bader}. The diffusion of the single Ca atom
adsorbed on the graphene has to overcome relatively small energy barriers of
$Q=$118 meV and 126 meV to diffuse to the top site (i.e. on top of C atom) and
bridge site (on top of the C-C bond), respectively. Ca atom adsorbed on the top or bridge sites
becomes less positively charged ($\sim0.89$ and $\sim0.92$ electrons,
respectively).

\begin{center}
\begin{figure}
\includegraphics[scale=0.35]{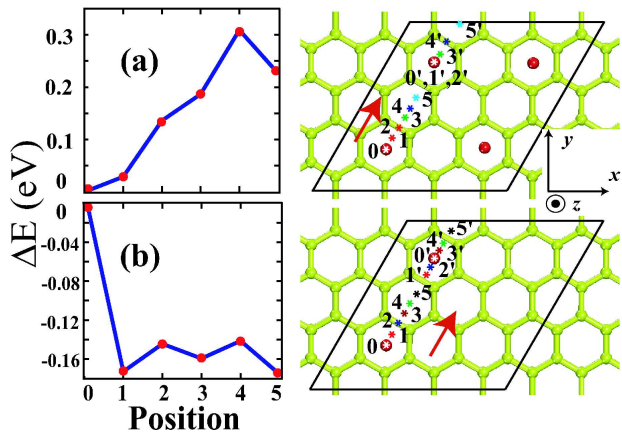}
\caption{(Color online) (a) Top-right panel: A (4$\times$4) cell of graphene having four Ca atoms. As Ca at the initial position 0 is moved in the direction of the arrow, its $z$-coordinate is optimized. The remaining three Ca atoms are fully relaxed. Beyond the position 2 of the first Ca, the Coulomb repulsion pushes the second Ca atom in the same direction through positions 3$^{'}$ 4$^{'}$ and 5$^{'}$ to maintain a distance with the first Ca. Top-left: The variation of energy as the first Ca is moves  through positions 1-5. (b) Bottom-right panel: Two Ca atoms adsorbed on each (4$\times$4) cell of graphene with their initial positions  0 and 0$^{'}$. As the first Ca moves from 0 to 1, the second one moves from 0$^{'}$ to 1$^{'}$ having the Ca-Ca distance of 3.74~\AA, whereby the energy is lowered by $\sim$0.176 meV. Two Ca atoms are prevented from being closer to each other and as the first Ca moves from 1 to 2,3,4 and 5 positions, the second one reverses his direction and moves through 2$^{'}$, 3$^{'}$, 4$^{'}$ and 5$^{'}$ in the same direction as the first Ca atom. Bottom-left: The variation of the energy with the positions of Ca atoms.}
\label{fig:stability}
\end{figure}
\end{center}

A denser Ca coverage, which is energetically more favorable, is attained,
if one Ca is adsorbed on each (2$\times$2) cell of graphene with Ca-Ca
distance of $4.92$ \AA. Ca atom adsorbed on the top and bridge sites has a
binding energy of $0.86$ and $0.89$ eV, respectively. However, energetically most
favorable adsorption site is found to be the H1 site, which is $2.15$ \AA~ above
the graphene with a binding energy of $1.14$ eV. Here, the Ca-Ca coupling is subtracted
from the calculated binding energy. In this dense (2$\times$2) coverage, a stronger electric
field is induced between Ca atoms and the graphene layer, which, in turn, leads to a
larger back-donation of charge from the graphene layer to $3d$-orbitals of Ca atom.
Hence by increasing Ca coverage from (4$\times$4) to (2$\times$2), adsorbed Ca atoms
become less positively charged, but their binding energy increases. As demonstrated in Fig.\ref{fig:stability}, even if it is energetically more favorable, the clustering of adsorbed Ca atoms is hindered by the Coulomb repulsion.

%Since the hydrogen storage capacity of the present Ca+graphene system could have been reduced by
%the clustering of adsorbed Ca atoms on graphene layer, we examined the possibility of clustering. Details
%are given in Fig.\ref{fig:stability}. Our results indicate that a stable and uniform Ca coverage
%on one sides of graphene up to $\Theta$=12.5 \% can be attained. The repulsive Coulomb interaction
%between Ca atoms hinders clustering. For example, single sided adsorption with $\Theta$=50 \%, where one Ca is adsorbed on each H1 sites on one side of graphene layer, is unstable and adsorbed Ca atoms repeal each other until there is only one adsorbed Ca on a single hollow site of the (2$\times$2) graphene unit cell.

\begin{center}
\begin{figure*}
\includegraphics[scale=0.45]{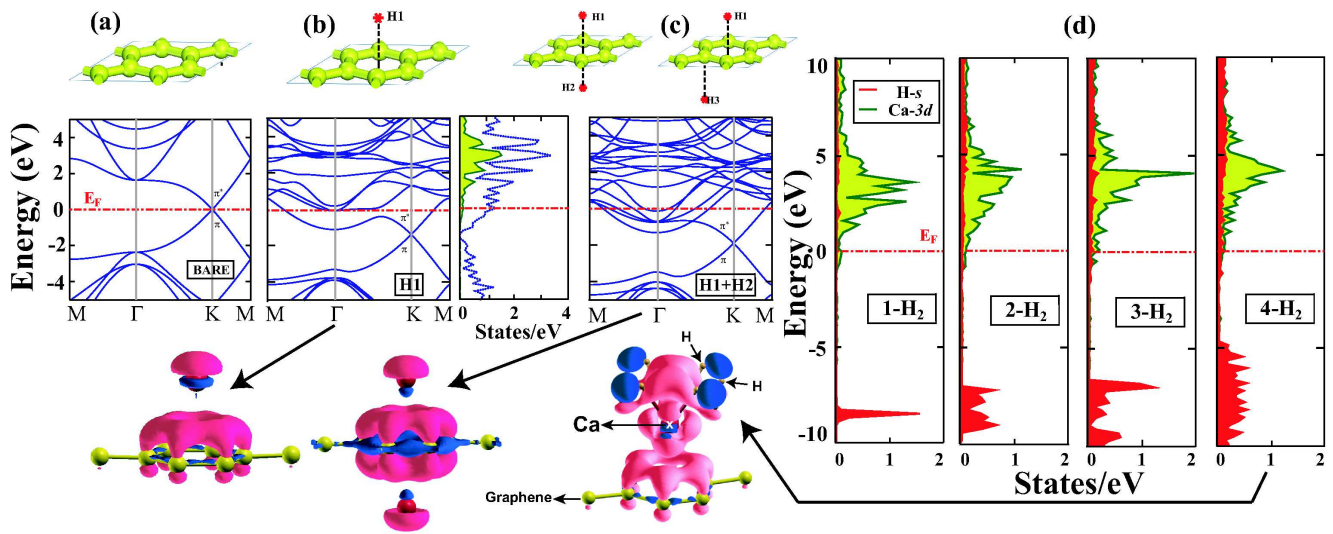}
\caption{(Color online) (a) The (2$\times$2) cell of graphene lattice and the energy band structure of bare graphene folded to the (2$\times$2) cell. (b) Single Ca atom is adsorbed on the H1 adsorption site of the (2$\times$2) cell of graphene, energy band structure and corresponding total density of states (dotted blue-dark curve) and partial density of states projected to Ca-\emph{3d} orbitals (green-gray). Isosurfaces of the difference charge density, $\Delta\rho$, with pink (light) and blue (dark) isosurfaces indicating charge accumulation and charge depletion regions. Isosurface charge density is taken to be 0.0038 electrons/\AA$^3$. (c) Similar to (b) (excluding the partial and total density of states), but Ca atoms are adsorbed on both sides of graphene at the H1 and H2 sites. (H1+H3 configuration is also shown.) (d) Partial densities of states on H-\emph{s} (red-dark) and Ca-\emph{3d} (green-gray) orbitals for 2, 3, and 4 H$_2$ absorbed in H1 configuration, and also isosurface of difference charge densities corresponding to 4H$_2$+Ca+Graphene configuration. Zero of band energy is set to the Fermi energy, E$_F$.}
\label{fig:band}
\end{figure*}
\end{center}

We next consider the double sided adsorption of Ca. The binding energy of second Ca atom for the double sided
adsorption with H1+H2 and H1+H3 configuration indicated in Fig.\ref{fig:band} (c), is $1.27$ and $1.26$
eV, respectively. Since the repulsive Coulomb interaction between Ca atoms on the upper and lower part of the plane is screened by the negative charge around graphene, the binding energy of Ca atom in the double sided adsorption is larger than that in the single sided adsorption. It is also found that $3d$-orbitals of both Ca atoms have higher occupancies as compared with Ca atom in the single sided adsorption. It is noted, however, that the partial occupancy of $3d$-orbitals of Ca atom does not cause any magnetic properties in the system. Our results indicate that a stable and uniform Ca coverage up to $\Theta$=12.5 \% ($\Theta$= 25 \%) can be attained for single sided (double sided with H1+H2 or H1+H3) adsorption forming a (2$\times$2) pattern.

Finite-temperature ab initio molecular dynamics simulations have
also been carried out for Ca adsorbed on the (2$\times$2) graphene unit cell for H1 geometry. Simulations are performed by normalizing the velocities of the ions and increasing the temperature of the system gradually from $0$ K to $900$ K in $300$ time steps. The duration of time steps are intentionally taken as $3$ fs, which is relatively longer for a MD calculation. If the system is unstable, the geometry of the structure can be destroyed much easier in long time steps. While the bonding between adsorbed Ca atom and the graphene layer is sustained, the adsorbed (2$\times$2) Ca layer begins to diffuse on the graphene layer as the temperature of the system rises over $\sim 300$ K. However, no structural deformation is observed indicating that the Ca+Graphene system is found to be stable up to $900$ K within 300 time steps.

Other alkaline-earth metals, such as beryllium and magnesium do not form strong bonds with graphene. Since Be has ionization potential of $9.32$ eV\cite{kittel} which is much higher than that of Ca atom ($6.11$ eV), the charge of its $2s$-orbital cannot easily transferred to the graphene layer. A similar situation occurs also with Mg having an ionization potential of $7.64$ eV. Besides, the hybridization of $\pi^{*}-$orbitals of graphene with the $d$- orbital of Ca atom, which is absent in both Be and Mg plays an essential role in strong binding of Ca to graphene. However, Ti and Co form strong bonds (with binding energies 1.58 and 1.20 eV for the (2$\times$2) adsorption pattern, respectively)\cite{sevincli}. The binding energies of Fe, Cr and Mo are rather weak.

The above arguments related with the binding of Ca to graphene are confirmed by examining the band structure and the charge difference isosurfaces presented in Fig.\ref{fig:band}. Both H1+H2 and H1+H3 adsorption configurations are included in our calculations because there is a small energy difference (H1+H2 structure is 26 meV more energetic.) between them. Hence, both adsorption configurations should be observable at room temperature conditions.  Charge difference isosurfaces are obtained by subtracting charge densities of Ca and bare graphene from that of Ca+graphene, namely $\Delta \rho = \rho_{Ca+Gr}-\rho_{Ca}-\rho_{Gr}$. It is seen that  there is a significant charge accumulation between the adsorbed Ca atom and graphene, which forms the ligand field. Partial occupation of $3d$-orbitals of Ca can be most clearly demonstrated by the projected density of states in Fig.\ref{fig:band} (b). The empty $\pi^*-$bands become occupied through charge transfer from $4s$-orbitals of adsorbed Ca and eventually get distorted due to $3d-\pi^*$ hybridization between $3d$-orbitals of Ca and the states of $\pi^*-$bands as a result of the charge back-donation process. Occupation of distorted graphene $\pi^*-$bands gives rise to the metallization of semi-metallic graphene sheets for all adsorption sites. It is also seen that charge density around graphene layer increased significantly as a result of double sided adsorption of Ca. The increase of charge back-donation to $3d$-orbitals becomes clear by the increased $3d$-projected density of states below the Fermi level. Changing the adsorption configuration from H1+H2 to H1+H3 does not make any essential changes in the electronic structure. One notes that the position of Fermi energy and hence electron density can be monitored by the controlled doping of Ca atoms. Metallization process is also important for graphene nanoribbons, which form conductive interconnects and spintronic devices in the same nanostructure\cite{sevincli,cohen}. It might be an interesting study to investigate the magnetic and electronic properties of Ca adsorption on graphene nanoribbons due to its different bonding mechanism.

\begin{center}
\begin{figure*}
\includegraphics[scale=0.65]{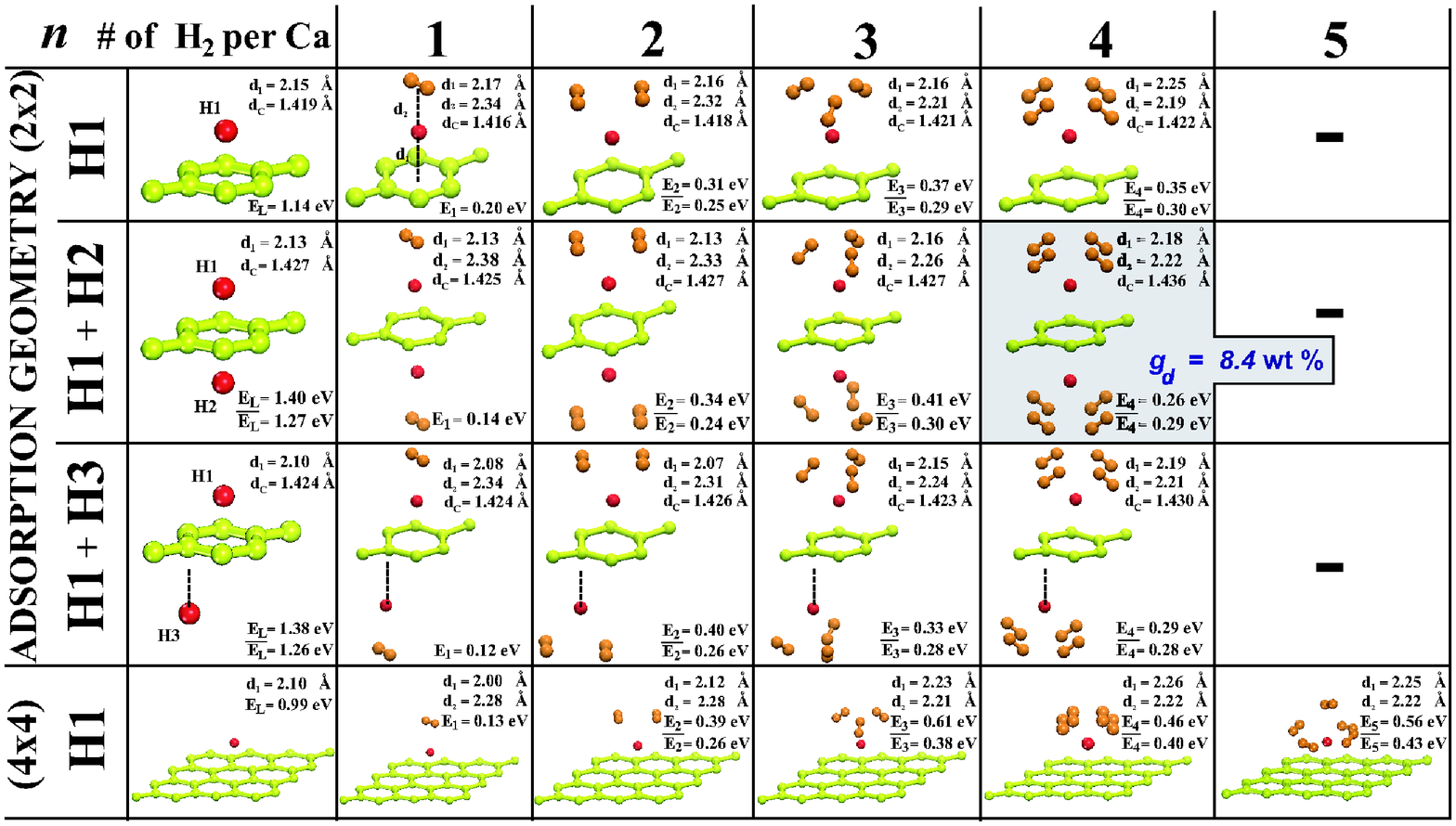}
\caption{(Color online) Sites and energetics of Ca
adsorbed on graphene with the (2$\times$2) coverage and absorption of
H$_2$ molecules by Ca atoms. $d_c$ is the average C-C distance in the graphene layer. $E_L$ is the binding energy of Ca
atom adsorbed on H1-site, which is a minimum energy site. For
H1+H2 or H1+H3 configurations corresponding to double sided
adsorption, $E_L$ is the binding energy of the second Ca atom and
$\overline E_L$ is the average binding energy. For H1, H1+H2 and
H1+H3 configurations, $E_1$ is the binding energy of the first
H$_2$ absorbed by each Ca atom; $E_n$ ($n$=2-5) is the binding
energy of the last $n^{th}$ H$_2$ molecule absorbed by each Ca
atom; $\overline E_n$ is the average binding energy of $n$ H$_2$
molecules absorbed by a Ca atom. Last row indicates the sites and energetics of one Ca atom adsorbed on each (4$\times$4) cell of graphene and absorption of H$_2$ molecules by each Ca atom. Only the (4$\times$4) coverage can absorb $5$ H$_2$ molecules. The shaded panel indicates energetically the most favorable H$_2$ absorption configuration.}
\label{fig:energy}
\end{figure*}
\end{center}

 We next study the absorption of hydrogen molecules by Ca atoms. A summary of energetics and geometry related with the absorption of molecular H$_2$ for H1, H1+H2, and H1+H3 sites for the (2$\times$2) and H1 site for the (4$\times$4) coverage are given in Fig.\ref{fig:energy}. The binding mechanism of H$_2$ invokes not only the adsorbed Ca atom, but also the graphene layer. In the case of single and double H$_2$ absorption, the absorbed molecules are parallel to graphene and all hydrogen atoms are equidistant from Ca atom. As a result, both hydrogen atom of each absorbed H$_2$ have the same excess charge of $\sim 0.08$ electrons. Once the number of H$_2$ absorbed by each Ca atom exceeded two, absorbed H$_2$ molecules tent to tilt towards Ca atom because of increased positive charge of Ca atom and the symmetry of the bonding configuration of H$_2$ molecules. The charge of Ca, H atom closer to Ca, H atom farther from Ca and graphene are calculated for 8H$_2$+2Ca+Graphene system corresponding to H1+H2 configuration in Fig.\ref{fig:energy} to be $\sim +1.29$, $\sim -0.06$, $\sim -0.11$ and $\sim -1.23$ electrons. One hydrogen atom of tilted H$_2$, which is closer to Ca has more excess charge than the other one. It is important to note that charges transferred to absorbed H$_2$ are not only from Ca atom. Graphene atoms at close proximity also supply charge through back-donation process. At the end, ionic bonding through attractive Coulomb interaction between positively charged Ca and negatively charged H and weak van der Waals interaction are responsible for the formation of mixed bonding between H$_2$ molecules and Ca adsorbed on graphene. The above discussion is substantiated by the partial density of states in Fig.\ref{fig:energy}(d). The excess charge on H-\emph{s} and Ca-\emph{3d} orbitals and their contribution to the states below the Fermi level increase with increasing number of H$_2$ molecules. Broadening of the molecular level of H$_2$ at $\sim -9$ eV indicates significant H$_2$-H$_2$ interaction, that in turn increases the binding energy. In fact, the binding energy of the first H$_2$ molecule to Ca atom which prefers to be parallel to the graphene layer is generally small. Whereas the average binding energy for two H$_2$ molecules which are again located parallel to the graphene layer, and for three or more H$_2$ molecules which are tilted around Ca atom are larger. We note that the adsorption of Ca atoms and also H$_2$ molecules slightly affect the underlying graphene lattice and C-C distance. The average C-C distance of bare graphene is increased from 1.41 \AA ~ upon adsorption of Ca and absorption of H$_2$ to $d_c$ values indicated in all (2$\times$2) structures in Fig.\ref{fig:energy}. Since Ca-Ca interaction is negligible in (4$\times$4) structures, there is no variation in average C-C distance.

Maximum number of absorbed H$_2$ per adsorbed Ca atom is four for the (2$\times$2) coverage yielding a H$_2$ storage capacity of $8.4$ wt \%  and five for  the (4$\times$4) coverage of graphene. The reason why we include the (4$\times$4) coverage even if the resulting gravimetric density is very low ($\sim\%2.3$ wt), is to mimic the Ca-H$_2$ interaction in the absence of H$_2$-H$_2$ interaction occurring in the (2$\times$2) coverage. Fifth H$_2$ molecule can be bound to the top of Ca atom in the (4$\times$4) coverage with a significantly high binding energy. Other $4$ H$_2$ molecules remain in quadrilateral positions around Ca. When we compare graphene with C$_{60}$\cite{yoon}, we can conclude that C$_{60}$ with a single Ca adsorbed on the surface yields similar results with the (4$\times$4) coverage on graphene. However, increasing Ca coverage adsorption results in lower binding energies of absorbed H$_2$ molecules in the present case. Unfortunately, we cannot comment on the case of high Ca coverage of C$_{60}$, since Yoon \emph{et al.}\cite{yoon} did not give details on the energetics of H$_2$ molecules in denser Ca adsorption. They have just emphasized that adsorption of $32$ Ca results in full coverage of C$_{60}$ surface and this structure can absorb up to $92$ H$_2$ molecules with binding energy of $\sim 0.4$ eV. Under these circumstances, single Ca atom can hold only 3 H$_2$ molecules. In graphene structures, while the charge on (ie. charge depletion or positive charge) Ca increases with increased number of the absorbed H$_2$ molecules, the electric field around Ca increases. This, in turn, results in a decrement in the distance between adsorbed Ca and polarized H$_2$ molecules. The charge on graphene decreases, as well.

In conclusion, this paper deals with two different subjects which are of current interest; namely graphene and hydrogen storage. Firstly, we showed that recently synthesized graphene with carriers behaving as if massless Dirac fermions can be metallized as a result of the adsorption of Ca atoms. Electrons donated by Ca is accommodated partly by the $\pi^*-$bands of graphene, partly back donated to its $3d$-orbitals. Ca atoms can be bound to both sides of graphene and can attain 25\% coverage without clustering. Secondly, we found that each adsorbed Ca can absorb up to four hydrogen molecules. At full coverage this yields a storage capacity of $\sim$8.4 wt\%, which is higher than the value set for feasible gravimetric density of hydrogen storage. The calculated bonding energies of hydrogen molecules are suitable for room temperature storage; while above the room temperature hydrogen molecules are released, Ca atoms remain adsorbed on graphene for further recycling. Even though storage capacities higher than the present case is achieved\cite{ataca} in different nanostructures, our results may be important for efficient hydrogen storage since graphene flakes are now easily available.

This work is supported by TUBITAK through the grant TBAG104536. Part of the computational resources for this study has been provided through Grant No. 2-024-2007 by the National Centre for High Performance Computing of Turkey, Istanbul Technical University.

\end{document}